\documentclass[a4paper,12pt]{article}
\usepackage{amssymb}
\usepackage{lscape}
\usepackage{tabls}
\def\beq{\begin{equation}}
\def\eeq{\end{equation}}
\def\bea{\begin{eqnarray}}
\def\eea{\end{eqnarray}}
\def\nn{\nonumber}
\def\h#1{\hat{#1}}

\def\U{{\cal U}}
\def\A{{\cal A}}

\def\eab{\exp_{\alpha, \beta}}
\def\nab{(n)_{\alpha, \beta}}
\def\Gab{\Gamma_{\alpha, \beta}}
\def\z2{|\widehat{\zeta}\rangle} 
\def\h{\U_{\langle q\rangle}(h(1))}
\def\H{{\cal F}un_{\langle q\rangle}(H(1))}

\def\oa{\mathfrak{a}}

\makeatletter
  \def\@cite#1#2{${\mbox{#1\if@tempswa , #2\fi}}$}
\makeatother
\def\ocite#1{$^{\mbox{\scriptsize{\cite{#1}}}}$}
\makeatletter
  \def\@biblabel#1{$^{\mbox{#1}}$}
\makeatother
\begin{document}
%
%
%
%
%
%
\thispagestyle{empty}

\vspace*{3cm}
\begin{center}
{\LARGE\sf
Generalized boson algebra and its entangled bipartite coherent states}

\bigskip\bigskip
N. Aizawa

\bigskip
\textit{
Department of Mathematics and Information Sciences, \\
Graduate School of Science,\\
Osaka Prefecture University, \\
Daisen Campus, Sakai, Osaka 590-0035, Japan}\\
\bigskip
R. Chakrabarti

\bigskip
\textit{
Department of Theoretical Physics, \\
University of Madras, \\
Guindy Campus, Chennai 600 025, India
}

\bigskip
J. Segar

\bigskip
\textit{
Department of Physics, \\
Ramakrishna Mission Vivekananda College, \\
Mylapore, Chennai 600 004, India
}
\bigskip

\end{center}

\vfill
\begin{abstract}
Starting with a given generalized boson algebra $\h$ known as the 
bosonized version of the quantum super-Hopf ${\cal U}_{q}(osp(1|2))$
algebra, we employ the Hopf duality arguments to provide the dually 
conjugate function algebra $\H$. Both the Hopf algebras being finitely 
generated, we produce a closed form expression of the universal 
${\cal T}$ matrix that caps the duality and generalizes the familiar 
exponential map relating a Lie algebra with its corresponding group. 
Subsequently, using an inverse Mellin transform approach, the 
coherent states of single-node systems subject to the $\h$ symmetry 
are found to be complete with a positive-definite integration measure. 
Nonclassical coalgebraic structure of the $\h$ algebra is found to 
generate naturally {\it entangled} coherent states in bipartite 
composite systems.   
\end{abstract}

\medskip

%
\newpage 
\setcounter{page}{1}
\section{Introduction}

Quantization of the boson algebra has been actively investigated 
\ocite{M89}${}^{-}$\ocite{APS01} due to its importance in studies 
of quantum groups, special functions, integrable models and the 
theory of noncommuting spaces. Many recent works in this 
area \ocite{Y91}${}^{-}$\ocite{MT97} focus on the 
quantized boson algebras endowed with Hopf structures as they are 
naturally equipped for applications in many-body systems of interest. 
In particular, it was observed by Macfarlane and Majid \ocite{MM92} 
that a quantum boson algebra admitting a Hopf structure plays the role 
of the spectrum generating algebra for the $q$-oscillator, 
\ocite{M89}${}^{-}$\ocite{SF89} as it is the bosonized version of the 
super-Hopf ${\cal U}_{q}(osp(1|2))$ algebra. This algebra was further 
generalized and studied in Refs. [\cite{PT97,TPJ97}]. These authors also 
pointed out the close relation of this algebra with the 
Calogero-Sutherland \ocite{C69, S71} type of models. In the present 
work, we study and make applications of this generalized boson algebra 
$\h$ defined in (\ref{U_alg}) and (\ref{U_coal}). 

\par

Using a technique developed by Fronsdal and Galindo \ocite{FG93} in the 
context of ${\cal U}_{q}(gl(2))$ algebra, we in Sec. \ref{Hopf_dual} 
study the Hopf duality and obtain the full Hopf structure of the 
function algebra $\H$, dually conjugate to the $\h$ algebra. The 
corresponding dual form, alternately referred to as the universal 
${\cal T}$ matrix, caps 
the duality structure and embodies the suitably modified exponential 
relationship $\h \rightarrow \H$. Noticing that both the Hopf algebras are 
finitely generated, we derive a closed form expression of the universal
${\cal T}$ matrix in terms of two sets of generators. The main 
usefulness of the universal ${\cal T}$ matrix stems from the fact that 
the transfer matrices of integrable models appear, upon specialization, 
in passing from operator structure to representations.     

\par

Enroute to our construction of the coherent states of the bipartite 
composite systems governed by $\h$ symmetry, we in 
Sec. \ref{completeness} provide a resolution of unity via the 
coherent states of the corresponding single-node systems. This property 
allows the coherent states to be complete (actually, overcomplete) set, 
and this is essential for a majority of applications in quantum mechanics. 
Recent works \ocite{SPS99}${}^{-}$\ocite{CV04} in establishing the 
resolution of the 
unit operator in an ensemble of generalized coherent states have used 
the method of inverse Mellin transform. Using an inverse Mellin 
transform of an associated Stieltjes moment problem, we obtain the 
resolution of unity of the single-node coherent states in the form of 
an ordinary integral with a positive-definite measure.

\par

Turning towards applications of the Hopf coalgebraic structure of 
the $\h$ algebra we note that it leads to qualitatively new properties 
of the many-body systems. In Sec. \ref{entanglement} we introduce and 
analytically obtain the coherent states in a bipartite composite 
system subject to $\h$ symmetry. The normalizable coherent states are 
{\it naturally entangled for a nonclassical value of} $q (\neq 1)$. 
The entanglement disappears in the classical $q \rightarrow 1$ limit.
Study of quantum information theory using entangled coherent states
is of much current interest. \ocite{S92, MMS00} Recently bipartite 
Barut-Girardello \ocite{BG71} coherent states of the 
${\cal U}_{q}(su(1,1))$ algebra have been found to be entangled
for $q \neq 1$. Our present calculation adds to the expectation that the 
{\it entanglement of bipartite and multipartite coherent states is a
generic feature of the quantum algebras}.
     
%
\section{The dual algebra and the universal ${\cal T}$ matrix}
\label{Hopf_dual}

Following the authors in Refs. [\cite{MM92, PT97, TPJ97}], we consider a 
$q$-deformed generalized boson algebra $\h$ generated by $a$, 
$a^{\dagger}$ and $N$ subject to the commutation relations
\beq
a a^{\dagger} + a^{\dagger} a = [\alpha N + \beta]_{q},\qquad
[N , a ]  =  - a, \qquad [N , a^{\dagger} ] = a^{\dagger}, 
\label{U_alg}
\eeq
where $q$ has generic real value, $(\alpha, \beta) \in \mathbb{R}$, and
$[{\cal X}]_{q} = \frac{q^{{\cal X}} - q^{-{\cal X}}}{q - q^{-1}}$.
The supplementary generating element ${\sf g} \equiv (-1)^{\tilde N}$, where 
${\tilde N} = N + \frac{\beta}{\alpha}\, {\mathbb  I}, \alpha \neq 0$, 
plays a key role in the construction of the Hopf coalgebraic structure. The 
coalgebraic maps read
\bea
& & \Delta(N) =   N \otimes {\mathbb I} + {\mathbb I} \otimes N 
+ \frac{\beta}{\alpha} {\mathbb I} \otimes {\mathbb I}, \quad 
\Delta( (-1)^{\tilde N} )=(-1)^{\tilde N} \otimes (-1)^{\tilde N},\nn \\
& & \Delta(a)= a  \otimes q^{\frac{\alpha {\tilde N} } {2}} 
+ (-1)^{\tilde{N}}\, q^{-\frac{\alpha {\tilde N} } {2}} \otimes a,\,\,\, 
\Delta(a^{\dagger}) = a^{\dagger}  \otimes q^{\frac{\alpha {\tilde N}}{2}} 
+ (-1)^{-\tilde N} q^{-\frac{\alpha {\tilde N} } {2}} \otimes a^{\dagger},
\nn\\
& &\epsilon(N) = - \frac{\beta}{\alpha}, \quad 
\epsilon( (-1)^{\tilde N} ) = 1 \quad 
\epsilon(a) = \epsilon(a^{\dagger}) =0,\nn\\ 
& & S(N) = - N - \frac{2 \beta}{\alpha},\quad 
S(a) = - (-1)^{- \tilde{N}}\,q^{- \alpha/2}\,a,\quad
S(a^{\dagger}) = a^{\dagger}\, (-1)^{\tilde{N}}\,q^{\alpha/2}.
\label{U_coal}
\eea
Imposing the constraint ${\sf g}^{2} = 1$, it has been found \ocite{MM92} 
that the algebra $\h$ is the bosonized version of the super-Hopf 
${\cal U}_{q}(osp(1|2))$ algebra. As it is endowed with bosonic statistical 
properties, $\h$ plays the role of the spectrum generating algebra 
\ocite{MM92} of the $q$-deformed oscillator. \ocite{M89}${}^{-}$\ocite{SF89} 
Following Refs. [\cite{PT97, TPJ97}], we do not impose the ${\sf g}^{2} = 1$
restriction. The universal ${\cal R}$ matrix of the $\h$ algebra has also 
been obtained. \ocite{MT97} 

\par

Two Hopf algebras $\U$ and $\A$ are in duality \ocite{FG93} if there 
exists a doubly-nondegenerate bilinear form
\beq
\langle , \rangle: ({\sf a}, {\sf u}) \rightarrow
\langle{\sf a}, {\sf u}\rangle \quad
\forall {\sf a} \in \A,\,\,\forall {\sf u} \in \U,
\label{au_form}
\eeq
such that, for $({\sf a}, {\sf b}) \in \A, 
({\sf u}, {\sf v}) \in \U$,
\bea
&&\langle {\sf a}, {\sf u v}\rangle = \langle \Delta_{\A}({\sf a}),
{\sf u} \otimes {\sf v}\rangle, \qquad
\langle {\sf a b}, {\sf u}\rangle = \langle {\sf a}\otimes {\sf b},
\Delta_{\U}({\sf u})\rangle,\nn\\
&&\langle {\sf a}, \mathbb{I}_{\U}\rangle = \epsilon_{\A}({\sf a}),\quad 
\langle \mathbb{I}_{\A}, {\sf u}\rangle = \epsilon_{\U}({\sf u}),\quad 
\langle {\sf a}, S_{\U}({\sf u})\rangle =  
\langle S_{\A}({\sf a}), {\sf u}\rangle.   
\label{dual_def}
\eea
Let the ordered monomials $E_{k \ell m} = a^{\dagger\, k} 
{\tilde N}^{\ell} a^{m},\,\, (k,\ell,m) \in (0, 1, 2,\cdots)$ be the 
basis elements of the $\h$ algebra obeying the multiplication and 
the induced coproduct rules given by 
\beq
E_{k \ell m} \,\, E_{k^{\prime} \ell^{\prime} m^{\prime}} = 
\sum_{p q r}\,
f^{p q r}_{k \ell m \,\, k^{\prime} \ell^{\prime} m^{\prime }} 
\,\, E_{p q r},\quad
\Delta( E_{k \ell m} ) = \sum_{p q r \atop p^{\prime} q^{\prime} r^{\prime}} 
g^{p q r \,\, p^{\prime}q^{\prime} r^{\prime}}_{k \ell m} \,\,
E_{p q r} \otimes E_{p^{\prime} q^{\prime} r^{\prime}}. 
\label{fg_def} 
\eeq
The basis elements $e^{k \ell m}$ of the dual Hopf algebra $\H$  
follows the relation
\beq
\langle e^{k \ell m}, E_{k^{\prime} \ell^{\prime} m^{\prime}} \rangle = 
\delta^{k}_{k^{\prime}}\,\delta^{\ell}_{\ell^{\prime}}\,
\delta^{m}_{m^{\prime}}.
\label{A_basis}
\eeq
In particular, the generating elements of the $\H$ algebra, defined as 
$x = e^{1 0 0}, y = e^{0 0 1}$ and $z = e^{0 1 0}$, satisfy the 
following duality structure:
\beq
\langle x, a^{\dagger}\rangle = 1, \,\,
\langle z, {\tilde N}\rangle = 1, \,\,
\langle y, a\rangle = 1. 
\label{xyz_def}
\eeq
The duality condition (\ref{dual_def}) requires the basis set $e^{k \ell m}$ 
to obey the multiplication and coproduct rules given below: 
\beq
e^{p q r}\,\,e^{p^{\prime} q^{\prime} r^{\prime}} = 
\sum_{k \ell m}\,
g^{p q r\,\,p^{\prime} q^{\prime} r^{\prime }}_{k \ell m}\, e^{k \ell m},
\qquad
\Delta(e^{pqr}) = \sum_{k \ell m \atop k^{\prime} \ell^{\prime} 
m^{\prime}} f^{p q r}_{k \ell m \,\, k^{\prime} \ell^{\prime} m^{\prime}} 
\,\,e^{k \ell m} \otimes e^{k^{\prime} \ell^{\prime} m^{\prime}}. 
\label{dual_struc} 
\eeq

\par

To derive the Hopf properties of the dual $\H$ algebra, we, therefore, 
need to extract the structure constants defined in (\ref{fg_def}).
Towards this end we note that the induced coproduct map of the 
elements $E_{k \ell m}$ may be obtained via (\ref{U_coal}):
\bea
\Delta(E_{k \ell m}) &=&  \Delta(a^{\dagger})^{k} \, 
\Delta({\tilde N}) ^{\ell}\, \Delta(a)^{m} \nn\\
&=& ( a^{\dagger}  \otimes q^{\alpha {\tilde N} / 2} 
+ \exp (-i \pi {\tilde N})\, q^{- \alpha {\tilde N} / 2 } \otimes 
a^{\dagger})^{k} \, ({\tilde N} \otimes I + I \otimes {\tilde N})^{\ell} 
\,\,\times \nn\\      
& & \times \,\,(a  \otimes q^{\alpha {\tilde N} / 2} 
+ \exp (i \pi {\tilde N})\, q^{- \alpha {\tilde N} / 2} \otimes a )^{m}, 
\label{E_copro}
\eea
where we have used $(- 1)^{\pm {\tilde N}} = \exp (\pm i \pi {\tilde N})$.
Employing ({\ref{E_copro}) we now obtain a set of structure constants: 
\bea
& & g^{1 0 0\,\,0 0 1}_{k \ell m} = \delta_{k 1} \, \delta_{\ell 0}\, 
\delta_{m 1}, \quad 
g^{0 0 1\,\,1 0 0}_{k \ell m} = \delta_{k 1}\,\delta_{\ell 0}\,\delta_{m 1}, 
\nn\\
& & g^{0 1 0\,\, 1 0 0}_{k \ell m} = \delta_{k 1} \, \delta_{\ell 1}\, 
\delta_{m 0} -
(\frac{\alpha}{2}\, \ln q + i \pi) \, \delta_{k 1} \, \delta_{\ell 0} \, 
\delta_{m 0}, \nn\\
& & g^{1 0 0\,\, 0 1 0}_{k \ell m} = \delta_{k 1} \, \delta_{\ell 1}\, 
\delta_{m 0} + 
\frac{\alpha}{2} \, \ln q \,\, \delta_{k 1} \, \delta_{\ell 0} \, 
\delta_{m 0}, \nn\\
& & g^{0 1 0\,\, 0 0 1}_{k \ell m} = \delta_{k 0} \, \delta_{\ell 1}\, 
\delta_{m 1} - (\frac{\alpha}{2}\, \ln q - i \pi)\, 
\delta_{k 0} \, \delta_{\ell 0} \, \delta _{m 1},\nn\\
& & g^{0 0 1\,\, 0 1 0}_{k\ell m} = \delta_{k 0} \, \delta_{\ell 1}\, 
\delta_{m 1} +  
\frac{\alpha}{2}\, \ln q \,\, \delta_{k 0} \, \delta_{\ell 0} \, 
\delta _{m 1}. 
\label{g_list}
\eea
The above structure constants immediately yield the algebraic 
relations obeyed by the generators of the $\H$ algebra:
\beq 
[x, y] = 0, \qquad [z, x] = - (\alpha \, \ln q + i \pi )\, x,  \qquad 
[z, y] = - (\alpha \, \ln q - i \pi )\, y. 
\label{xyz_alg}
\eeq
A representation of the above Lie algebra with complex structure 
constants may be easily obtained in terms of harmonic oscillators 
$\{\oa_{i}, \oa^{\dagger}_{i}\, |\, i = (1, 2)\}$ obeying the algebra
$[\oa_{i}, \oa_{j}^{\dagger}] = \delta_{i\,j},\,[\oa_{i}, \oa_{j}] = 0, \,
[\oa_{i}^{\dagger}, \oa_{j}^{\dagger}] = 0,\,(i, j) = (1, 2)$:  
\beq
x = \oa_{1}, \quad y = \oa_{2}, \quad
z = \alpha\, \ln q \,(\oa_{1}^{\dagger}\,\oa_{1} 
+ \oa_{2}^{\dagger}\,\oa_{2}) + i \pi\, 
(\oa_{1}^{\dagger}\,\oa_{1} - \oa_{2}^{\dagger}\,\oa_{2}). 
\label{xyz_rep}
\eeq

\par

Proceeding towards constructing the coproduct maps of the 
generating elements of the dual $\H$ algebra we notice that the defining 
properties (\ref{dual_struc}) provide the necessary recipe:
\bea
\Delta(x)&=& \sum_{k \ell m \, \atop k^{\prime} \ell^{\prime} m^{\prime}}
\, f^{1 0 0}_{k \ell m\; k^{\prime} \ell^{\prime} m^{\prime}} \, 
e^{k \ell m} \, \otimes 
e^{k^{\prime} \ell^{\prime} m^{\prime}},\nn\\
\Delta(z)&=& \sum_{k \ell m \, \atop k^{\prime} \ell^{\prime} m^{\prime}}\, 
f^{0 1 0}_{k \ell m\; k^{\prime} \ell^{\prime} m^{\prime}}  \,  
e^{k \ell m} \, \otimes e^{k^{\prime} \ell^{\prime} m^{\prime}}, \nn\\
\Delta(y)&=& \sum_{k \ell m \, \atop k^{\prime} \ell^{\prime} m^{\prime}}\, 
f^{0 0 1}_{k \ell m\; k^{\prime} \ell^{\prime} m^{\prime}} 
\, e^{k \ell m} \, \otimes e^{k^{\prime} \ell^{\prime} m^{\prime}}. 
\label{xyz_copro}
\eea
The relevant structure constants obtained via (\ref{fg_def}) are listed 
below: 
\bea
&&f^{1 0 0}_{k \ell m\, k^{\prime} \ell^{\prime} m^{\prime}} = 
\delta_{k 1}\, \delta_{\ell 0}\, \delta_{m 0}\, \delta_{k^{\prime} 0}\,  
\delta_{\ell^{\prime} 0}\, \delta_{m^{\prime } 0} + {\sigma}_{m+1}\,\, 
\delta_{k 0}\, \delta_{k^{\prime}\, m+1} \, \delta_{\ell^{\prime} 0} \, 
\delta_{m^{\prime } 0},\nn\\
&&f^{0 1 0}_{k \ell m\, k^{\prime} \ell^{\prime} m^{\prime}} = 
\delta_{k 0}\, \delta_{m 0}\,\delta_{k^{\prime} 0}\, 
\delta_{m^{\prime} 0}\,(\delta_{\ell\, 1}\, \delta_{\ell^{\prime}\, 0} + 
\delta_{\ell\, 0}\, \delta_{\ell^{\prime}\, 1})\, + \, 
\frac {2 \alpha \,\ln q}{q - q^{-1}}\,\, {\sigma}_{m}\,\, 
\delta_{k 0}\, \delta_{\ell 0} \, \delta_{k^{\prime} m} \, 
\delta_{\ell^{\prime} 0} \delta_{m^{\prime} 0},\nn\\
&&f^{0 0 1}_{k \ell m\,k^{\prime} l^{\prime} m^{\prime}} = 
\delta_{k 0}\, \delta_{\ell 0}\, \delta_{m 0}\,\delta_{k^{\prime} 0}\, 
\delta_{\ell^{\prime} 0}\, \delta_{m^{\prime} 1} +  
{\sigma}_{k^{\prime} + 1}\, \delta_{k 0}\, \delta_{\ell 0} \, 
\delta_{m\, k^{\prime} + 1} \, \delta_{m^{\prime} 0},\nn\\ 
&&\sigma_{1} = 1,\qquad
\sigma_{m\,(> 1)} = \prod_{k = 1}^{m - 1}\,\sum_{\ell = 0}^{k - 1}\,
(- 1)^{\ell}\,[(k - \ell) \alpha]_{q}.
\label{f_copro}
\eea
The coproduct maps of the dual generators may now be explicitly 
obtained {\it {\`a} la} (\ref{xyz_copro}) provided the basis elements
$e^{k \ell m}$ of the dual $\H$ algebra are known. We complete this 
task subsequently. 

\par 

As the dual algebra $\H$ is finitely generated, we may start with the 
generators $(x, y, z)$ and obtain all dual basis elements
$e^{k \ell m},\;(k, \ell, m) \in (0, 1, 2,\cdots)$ by successively 
applying the multiplication rule given in the first equation in 
(\ref{dual_struc}). The necessary structure constants may be read 
from the relation (\ref{fg_def}) of the $\h$ algebra. In the procedure 
described below we maintain the operator ordering of the monomials as 
$x^{k} z^{\ell} y^{m},\,\, (k, \ell, m) \in (0, 1, 2,\cdots)$. The 
product rule 
\beq
e^{1 0 0} \,\, e^{p 0 0} = \sum_{k \ell m}\, 
g^{1 0 0\,p 0 0}_{k \ell m}\, e^{k \ell m}
\label{prod_1} 
\eeq
and the explicit evaluation of the structure constant
\beq
g^{1 0 0\,p 0 0}_{k \ell m} = \{k\}_{q^\alpha} \, \delta_{k\;p+1} \, 
\delta_{\ell 0} \, \delta_{m 0},\qquad 
\{n\}_{q} = \frac{q^{n/2} - (-1)^{n} q^{- n/2}}{q^{1/2} + q^{- 1/2}}
\label{g_1}
\eeq
obtained from the second equation in (\ref {fg_def}) immediately provide
\beq
e^{k 0 0} = \frac{x^k}{\{k\}_{q^{\alpha}}!}, \qquad
\{n\}_{q}! = \prod_{\ell=1}^{n} \{\ell\}_{q},\,\,\{0\}_{q}! = 1. 
\label{e_1}
\eeq
Employing another product rule
\beq
e^{p r 0} \, e^{0 1 0} = \sum_{k \ell m} g^{p r 0 \, 0 1 0}_{k \ell m} 
\,\, e^{k \ell m}
\label{prod_2}
\eeq   
and the value of the relevant structure constant 
\beq
g^{p r 0\,0 1 0}_{k \ell m} = (r + 1)\, \delta_{k p} \, 
\delta_{\ell \, r+1} \, \delta_{m 0} + \frac{\alpha p}{2}\,\, \ln q \,\, 
\delta_{k p}\, \delta_{\ell r}\, \delta_{m 0}
\label{g_2}
\eeq
obtained in the aforesaid way we produce the following result:
\beq
e^{k \ell 0} = \frac{x^k}{\{k\}_{q^{\alpha}}!}\,\,
\frac{\Big(z - \frac{\alpha}{2}\,\, k \,\ln q\Big)^{\ell}}{\ell!}.
\label{e_2}
\eeq
Continuing the above process of building of the dual basis set we use 
the product rule
\beq
e^{p r s} \, e^{0 0 1} = \sum_{k \ell m} g^{p r s \, 0 0 1}_{k \ell m} 
\,\, e^{k \ell m}
\label{prod_3}
\eeq   
and the value of the corresponding structure constant
\beq
g^{p r s\,0 0 1}_{k \ell m} = \{m\}_{q^\alpha} \, 
\sum_{j = 0}^{r}\,\frac{1}{j!}\,
\Big(- \frac{\alpha}{2}\,\ln q + i \pi\Big)^{j}\,\,
\delta_{k p} \, 
\delta_{\ell \; r - j} \, \delta_{m  \; s + 1} 
\label{g_3}
\eeq
obtained via (\ref{fg_def}). This finally leads us to the complete 
construction of the dual basis element:
\beq
e^{k \ell m} = \frac{x^k}{\{k\}_{q^{\alpha}}!}\,\,
\frac{\Big(z - \frac{\alpha}{2}\,\, (k - m) \,\ln q \, - 
i m \pi\Big)^{\ell}}{\ell!} 
\frac{y^m}{\{m\}_{q^{\alpha}}!}.
\label{e_3}
\eeq
Combining our results in (\ref{xyz_copro}), (\ref{f_copro}) and 
(\ref{e_3}), we now provide the promised coproduct structure of the 
generators of the $\H$ algebra:
\bea
&&\Delta(x) = x \otimes {\mathbb I} + \sum_{m = 0}^{\infty}\, 
(- 1)^{m}\,q^{m \alpha/ 2}\, \sigma_{m + 1}\, \exp (z)\,\,
\frac{y^{m}}{\{m\}_{q^{\alpha}}!} \otimes
\frac{x^{m + 1}}{\{m + 1\}_{q^{\alpha}}!},\nn\\
&&\Delta(z) = z \otimes {\mathbb I} + {\mathbb I} \otimes z + 
\frac{2 \alpha\,\, \ln q}{q - q^{- 1}}\,\, 
\sum_{m = 1}^{\infty}\, \sigma_{m}\,
\frac{y^m}{\{m\}_{q^{\alpha}}!} \otimes
\frac{x^{m}}{\{m\}_{q^{\alpha}}!},\nn\\
&&\Delta(y) = {\mathbb I} \otimes y + \sum_{m = 0}^{\infty}\, 
q^{- m \alpha/ 2}\, \sigma_{m + 1}\, 
\frac{y^{m + 1}}{\{m + 1\}_{q^{\alpha}}!} \otimes
\frac{x^{m}}{\{m\}_{q^{\alpha}}!}\,\exp (z).
\label{xyz_copro_1}
\eea 
Algebraic simplifications allow us to express the 
coproduct maps of the above generators more succinctly:
\bea
&&\Delta(x) = x \otimes {\mathbb I} + \sum_{m = 0}^{\infty}\, (- 
1)^{m}\,
\Big(\frac{q^{\alpha} + 1}{q - q^{-1}}\Big)^{m}\, \exp (z)\,\, 
y^{m}\otimes x^{m + 1}, \nn\\
&&\Delta(z) = z \otimes {\mathbb I} + {\mathbb I} \otimes z + 
\frac{2 \alpha\,\, \ln q}{q - q^{- 1}}\,\, 
\sum_{m = 1}^{\infty}\, \frac{1}{\{m\}_{q^{\alpha}}}
\Big(\frac{q^{\alpha/ 2} + q^{ - \alpha /2}}{q - q^{-1}}\Big)^{m - 1}
\,\, y^{m} \otimes x^{m},\nn\\ 
&&\Delta(y) = {\mathbb I} \otimes y + \sum_{m = 0}^{\infty}\, 
\Big(\frac{1 + q^{- \alpha}}{q - q^{- 1}}\Big)^{m}\,  
y^{m + 1} \otimes x^{m} \, \exp (z).
\label{xyz_copro_2}
\eea 
With the aid of the result (\ref{xyz_copro_2}) we may explicitly 
demonstrate that the coproduct map is a homomorphism of the algebra
(\ref{xyz_alg}): namely,
\bea
&&[\Delta(x), \Delta(y)] = 0, \qquad 
[\Delta(z), \Delta(x)] = - (\alpha \, \ln q + i \pi )\, \Delta(x),\nn\\ 
&&[\Delta(z), \Delta(y)] = - (\alpha \, \ln q - i \pi )\, \Delta(y). 
\label{alg_homo}
\eea
The coassociativity constraint 
\beq
(\hbox{id} \otimes \Delta) \circ \Delta({\cal X}) =     
(\Delta \otimes \hbox{id}) \circ \Delta({\cal X}) \qquad 
\forall {\cal X} \in (x, y, z) 
\label{xyz_coaas}   
\eeq
may also be established by using the following identity:
\bea
&& \exp (\Delta (z)) = (\exp (z) \otimes {\mathbb I})\,\, 
\prod_{m = 1}^{\infty}\,{\cal P}_{m}\,\,({\mathbb I} \otimes 
\exp (z))\nn\\
&& {\cal P}_{m} = \exp \left((- 1)^{m}\,
\frac{[m \, \alpha]_{q}}{m\,\{m\}_{q^{\alpha}}}\,\,
\Big(\frac{q^{\alpha/2} + q^{- \alpha/2}}{q - q^{- 1}}\right)^{m -1}\,\,
y^{m} \otimes x^{m}\Big).
\label{exp_del_z}
\eea   

\par

The counit map of the generators of the $\H$ algebra reads as
\beq
\epsilon (x) = \epsilon (y) = \epsilon (z) = 0.
\label{xyz_counit}
\eeq
The antipode map of the dual generators follows from the 
last equation in (\ref{dual_def}). We quote the results here: 
\bea
&& S(x) = \sum_{m = 0}^{\infty}\,(- 1)^{m}\,q^{(m + 2)\,\alpha/ 2}\,
\sigma_{m + 1}\, \frac{x^{m + 1}}{\{m + 1\}_{q^{\alpha}}!}\,
\exp(- (m + 1) z)\,\frac{y^{m}}{\{m\}_{q^{\alpha}}!},\nn\\
&& S(z) = - z + \frac{2 \alpha\, \ln q}{q - q^{-1}}\,
\sum_{m = 1}^{\infty}\, \sigma_{m}\,
\frac{x^{m}}{\{m\}_{q^{\alpha}}!}\,
\exp(- m z)\,\frac{y^{m}}{\{m\}_{q^{\alpha}}!},\nn\\
&& S(y) = \sum_{m = 0}^{\infty}\,q^{- (m + 2)\,\alpha/ 2}\,
\sigma_{m + 1}\, \frac{x^{m}}{\{m\}_{q^{\alpha}}!}\,
\exp(- (m + 1) z)\,\frac{y^{m + 1}}{\{m + 1\}_{q^{\alpha}}!}.
\label{xyz_antipode}
\eea
In an order by order calculation we may verify that the above antipode 
map is an antihomomorphism of the algebra (\ref{xyz_alg}), and the  
necessary Hopf constraint holds:
\beq
m \circ (S \otimes \hbox{id}) \circ \Delta({\cal X}) = 
m \circ (\hbox{id} \otimes S) \circ \Delta({\cal X}) =
\epsilon ({\cal X})\,{\mathbb I} \qquad       
\forall {\cal X} \in (x, y, z), 
\label{anti_cons}
\eeq   
where $m$ is the multiplication map. This completes our construction of 
the Hopf algebra $\H$ dually related to the generalized boson algebra 
$\h$.

\par

Our explicit listing of the complete set of dual basis elements in 
(\ref{e_3}) allows us to obtain {\it \`{a} la} Fronsdal and 
Galindo \ocite {FG93} the universal ${\cal T}$ matrix:
\beq
{\cal T} = \sum_{k, \ell, m}\, e^{k \ell m} \otimes E_{k \ell m}
\equiv {\cal T}_{e, E}.
\label{T_def}
\eeq 
The notion of the universal ${\cal T}$ matrix is a key feature capping 
the Hopf duality structure. Consequently, the duality relations 
(\ref{dual_def}) may be concisely expressed \ocite{FG93} in terms of the 
${\cal T}$ matrix as
\bea
&&{\cal T}_{e, E}\; {\cal T}_{e^{\prime}, E} = {\cal T}_{\Delta (e), E},
\qquad{\cal T}_{e, E}\; {\cal T}_{e, E^{\prime}} = 
{\cal T}_{e, \Delta (E)},\nn\\
&&{\cal T}_{\epsilon (e), E} = {\mathbb I}, \quad
{\cal T}_{e, \epsilon (E)} = {\mathbb I}, \quad
{\cal T}_{S(e), E} = {\cal T}_{e, S(E)},
\label{T_cap}
\eea
where $e$ and $e^{\prime}$\, ($E$ and $E^{\prime}$) refer to the two 
identical copies of $\H$ ($\h$) algebra.

\par 

As both the Hopf algebras in our case are finitely generated the 
universal ${\cal T}$ matrix may now be obtained as an operator valued 
function in a closed form: 
\beq
{\cal T} = {}_{\times}^{\times} \,{\cal E}\hbox{xp}_{q^{\alpha}}
\Big(x \otimes a^{\dagger}\, q^{ - \alpha\, {\tilde N}/ 2}\Big)\,
\exp (z \otimes {\tilde N})\, {\cal E}\hbox{xp}_{q^{\alpha}}
\Big(y \otimes (- 1)^{- \tilde N} \, q^{\alpha\, {\tilde N}/ 2}\, a\Big)\,
{}_{\times}^{\times},   
\label{univ_T} 
\eeq  
where ${\cal E}\hbox{xp}_{q} ({\cal X}) = \sum_{m = 0}^{\infty}\,
\frac{{\cal X}^{n}}{\{n\}_{q}!}$. The operator ordering has been 
explicitly indicated above. The universal ${\cal T}$ matrix, as evidenced
in (\ref{univ_T}), may be viewed \ocite{FG93} as the appropriate quantum 
group generalization of the familiar exponential map relating 
a Lie algebra with the corresponding Lie group. We note that the 
deformed exponential in (\ref{univ_T}) is different from that in Ref. 
[\cite{FG93}]. The universal ${\cal T}$ matrix given in (\ref{univ_T}) is 
endowed with a group-like coproduct rule and it is characterized by 
{\it noncommuting} parameters $(x, y, z)$ in a representation-independent 
way.

\section{Coherent states of a single-node system and their completeness}
\label{completeness}
\setcounter{equation}{0}
As a prelude to our subsequent construction of the entangled coherent 
states in a bipartite composite system, we, in the present section, 
study the completeness of the coherent states in a single-node system 
possessing the deformed Heisenberg symmetry $\h$ defined in 
(\ref{U_alg}) and (\ref{U_coal}). A Fock-type representation of 
the algebra $\h$ is given by \ocite{PT97, TPJ97}
\bea
& & |n\rangle = \frac{1}{\sqrt{(n)_{\alpha, \beta}!}}\,(a^{\dagger})^{n}
\,|0\rangle \quad n \in (0, 1, 2, \cdots),\qquad a |0 \rangle = 0,\nn\\
& & a\, |n\rangle = \sqrt{(n)_{\alpha, \beta}}\,| n-1\rangle, \quad
a^{\dagger}\,|n\rangle = \sqrt{(n+1)_{\alpha, \beta}}\,| n+1\rangle, 
\quad (-1)^{N}\,|n\rangle = (-1)^{n}\,|n\rangle,\nn \\
& & (n)_{\alpha, \beta} = (q^{\alpha / 2} + q^{ - \alpha /2})^{-1}\,
([n \alpha + \beta - \alpha / 2]_{q} + (-1)^{n + 1} \, 
[\beta - \alpha / 2]_{q}),\nn \\
& & (n)_{\alpha, \beta}! = \prod_{\ell = 1}^{n}\,(\ell)_{\alpha, \beta},
\quad (0)_{\alpha, \beta}! = 1, \quad 
\langle n | n^{\prime} \rangle = \delta_{n\, n^{\prime}}.
\label{Fock_state}
\eea  
For a single-node system the coherent state is defined as 
\beq
a |\zeta\rangle = \zeta |\zeta\rangle, \quad \zeta \in \mathbb{C}.    
\label{coh_def}
\eeq
The normalized coherent state reads
\beq
|\zeta\rangle = \frac{1}{\sqrt{\eab (|\zeta|^{2})}}\,
\sum_{n = 0} ^{\infty}\, \frac{\zeta^{n}}
{\sqrt{(n)_{\alpha, \beta}!}}\, |n\rangle 
= \frac{1}{\sqrt{\eab (|\zeta|^{2})}}\,
\eab (\zeta a^{\dagger}) |0\rangle,
\label{coh_const}
\eeq
where the deformed exponential is given by 
\beq
\eab ({\cal X}) = \sum _{n=0}^{\infty}\frac{{\cal X}^{n}}{\nab !}.
\label{def_exp}
\eeq
The coherent states (\ref{coh_const}) have nonvanishing inner products, 
and, therefore, are not orthogonal:
\beq
\langle \zeta^{\prime} | \zeta \rangle = 
\frac{\eab (\overline{\zeta^{\prime}} \zeta)}
{\sqrt{\eab (|\zeta^{\prime}|^{2})\,\eab (|\zeta|^{2})}}.
\label{inn_prod}
\eeq 

\par

Assuming the completeness of the discrete basis states 
\beq
\sum _{n = 0}^{\infty} \,|n\rangle \, \langle n| = {\mathbb I},
\label{n_com}
\eeq 
we now prove that the coherent states $|\zeta\rangle$ possess a 
resolution of identity with a positive definite integration measure in 
the complex plane. For constructing this measure we proceed by defining 
a generalized Gamma function suited to our purpose:
\beq
\Gab (z) = \prod _{\ell = 1}^{\infty}\, \frac{(\ell)_{\alpha, \beta}}
{(z + \ell -1)_{\alpha, \beta}}, \,\,\,
\Gab (z + 1) = (z)_{\alpha, \beta}\,\Gab (z),\,\,\,
\Gab (n + 1) = \nab !.
\label{def_Gam}
\eeq
In the limit $\alpha \rightarrow 2, \beta \rightarrow 1, q \rightarrow 
1$, the deformed Gamma function  $\Gab (z)$ reduces to its classical 
partner $\Gamma (z)$. Analytic continuation of $(n)_{\alpha, \beta}$ 
defined in (\ref{Fock_state}) for noninteger arguments may be done in 
two possible ways $(n \rightarrow z\,\, \Rightarrow\,\, 
(-1)^{n} \rightarrow \exp (\pm i \pi z))$ yielding results related to 
each other by complex conjugation:
\beq
(z)_{\alpha, \beta}^{(\pm)} = (q^{\alpha / 2} + q^{ - \alpha /2})^{-1}\,
([z \alpha + \beta - \alpha / 2]_{q} - \exp (\pm i \pi z) \, 
[\beta - \alpha / 2]_{q}).
\label{z_anacon}
\eeq 
The generalized $\Gab (z)$ functions (\ref{def_Gam})
corresponding to the said two analytic continuations are referred to as 
$\Gab^{(\pm)} (z)$. Omitting the superscripts here, we note that the 
singularity structure of the generalized Gamma function may be derived
from the following iterated relation:
\beq
\Gab (z) = \frac{\Gab (z + n + 1)}{(z) _{\alpha, \beta}\, 
(z + 1)_{\alpha, \beta}\,\cdots\, (z + n)_{\alpha, \beta}}.
\label{gam_iter}
\eeq       
The two analytic continuations given in (\ref{z_anacon}), in the limit 
$\varepsilon \rightarrow 0$, yield
\beq
(\varepsilon)_{\alpha, \beta}^{(\pm)} = \varepsilon \Big(\alpha\, 
[[\beta - \alpha/2]]\, \frac{\ln q}{q - q^{-1}} \mp i \,\varpi \,
[\beta - \alpha/2]_{q}\Big),
\label{epsi_lim}
\eeq
where $[[{\cal X}]] = \frac{q^{\cal X} + q^{ - \cal X}}
{q^{\alpha/2} + q^{- \alpha/2}}, 
\varpi = \frac{\pi}{q^{\alpha/2} + q^{- \alpha/2}}$. Using 
(\ref{gam_iter}) and (\ref{epsi_lim}) the singularity structure of the 
conjugate functions $\Gab^{(\pm)}(z)$ in the neighbourhood 
$z = - n + \varepsilon$ may be obtained as 
\beq
\Gab^{(\pm)}(-n + \varepsilon) = 
\frac{1}{(\varepsilon)^{(\pm)}_{\alpha, \beta}\,\,
(n)_{- \alpha, \beta - \alpha}!},        
\label{gam_sing}
\eeq
where we have used $(- n)_{\alpha, \beta} = 
(n)_{- \alpha, \beta - \alpha}$. As we are interested in the 
positive definiteness of the integration measure the said analytic 
continuation must be done in a {\it symmetric way by taking an average 
of the two complex conjugate functions}:
\beq
\Gab^{sym} (z) = (\Gab^{(+)} (z) + \Gab^{(-)} (z))/ 2.
\label{g_sym}
\eeq 
The singularity of the above symmetrized deformed Gamma function 
is obtained by using (\ref{gam_sing}):
\bea
\Gab^{sym} (-n + \varepsilon) &=& (\Gab^{(+)}(-n + \varepsilon) + 
\Gab^{(-)}(-n + \varepsilon))/2 \nn\\  
&=& \varepsilon^{-1}\,\frac{\cal P}{(n)_{- \alpha, \beta - \alpha}!},
\label{ave_sing}
\eea
where
\beq
{\cal P} = \frac{\alpha \,\, [[\beta - \alpha/2]]\,\,\ln q/(q - q^{-1})} 
{(\alpha\,\, [[\beta - \alpha/2]]\,\,\ln q/ (q - q^{-1}))^{2} + 
\varpi^{2}\,\,[\beta - \alpha/2]_{q}^{2}}.
\label{res_val}
\eeq
Parallel to the undeformed Gamma function, our $\Gab^{sym}(z)$ 
also possess, as evident from above, simple poles at 
$z = 0, -1, -2, \cdots$. Keeping in mind the above singularity structure 
of the generalized $\Gab^{sym}(z)$, we now obtain a resolution of 
the identity via coherent states $|\zeta \rangle$ in the form
\beq
\int \hbox{d}\mu(\zeta)\,\,\, |\zeta\rangle \langle\zeta| = \mathbb{I},
\label{res_unit}
\eeq
where the integration measure $\hbox{d}\mu(\zeta)$ is determined below. 
Using the polar decomposition $\zeta = \rho \exp(i \theta)$ with our 
construction of the coherent state (\ref{coh_const}), we integrate the 
angular variable $\theta$ to obtain
\beq
\eab(\rho^{2}) \, \int_{0}^{2 \pi}\, \frac{\hbox{d}\theta}{2 \pi}\,\,
|\zeta\rangle \langle\zeta| = \sum_{n = 0}^{\infty}\,\frac{\rho^{2 n}}
{\nab !}\,\,|n\rangle \langle n|. 
\label{ang_int}
\eeq   
Multiplying both sides of the above equation by a yet to be determined 
function $F(\rho)$, and integrating over the entire complex $\zeta$ 
plane, we get
\beq
\int\hbox{d}^{2}\zeta\,\eab(|\zeta|^{2})\, F(|\zeta|) 
|\zeta\rangle \langle\zeta| = \sum_{n = 0}^{\infty}\,\frac{{\cal I}_{n}}
{\nab !}\,\,|n\rangle \langle n|, 
\label{F_intro}
\eeq   
where $\hbox{d}^{2}\zeta = (2 \pi)^{-1}\,\rho\, \hbox {d}\rho 
\hbox {d}\theta$, and ${\cal I}_{n}$ represents the Mellin 
transform of the function $F(\rho)$:
\beq
{\cal I}_{n} = \int_{0}^{\infty}\hbox {d}\rho\, \rho^{2n + 1}\,F(\rho).
\label{In_def}
\eeq
If we now choose the transform ${\cal I}_{n}$ in (\ref{F_intro}) as
\beq
{\cal I}_{n} = \nab !,
\label{In_choice}
\eeq
it immediately follows that by the virtue of completeness relation 
(\ref{n_com}) of the discrete basis states $|n\rangle$, the rhs in 
(\ref{F_intro}) reduces to identity operator:
\beq
\int\hbox{d}^{2}\zeta\,\eab(|\zeta|^{2})\, F(|\zeta|) 
|\zeta\rangle \langle\zeta| = \mathbb{I}.
\label{F_invert}
\eeq   
The function $F(\rho)$ defined by the Stieltjes moment relation may now 
be explicitly obtained in terms of an inverse Mellin transform as
\beq
F(\rho) = \frac{1}{\pi i}\,\int_{c - i \infty}^{c + i \infty}\,
\hbox{d}z\, \rho^{- 2 z}\, (z - 1)_{\alpha, \beta}! 
= \frac{1}{\pi i}\,\int_{c - i \infty}^{c + i \infty}\,
\hbox{d}z\, \rho^{- 2 z}\, \Gab^{sym}(z).
\label{inv_Mel}
\eeq
In the second equation we have used, as explained earlier in the context 
of (\ref{g_sym}), a symmetrized analytic continuation of the deformed 
factorial. Using the singularity structure (\ref{ave_sing}) we now 
explicitly evaluate the previously undetermined measure function $F(\rho)$ 
via the contour integral (\ref{inv_Mel}) as the integral vanishes 
exponentially as $|z| \rightarrow \infty$ on the left-half plane: 
\beq
F(|\zeta|) = \frac{2 \alpha \,\, [[\beta - \alpha/2]]\,\,
\ln q/(q - q^{-1})} 
{(\alpha\,\, [[\beta - \alpha/2]]\,\,\ln q/ (q - q^{-1}))^{2} + 
\varpi^{2}\,\,[\beta - \alpha/2]_{q}^{2}}\,\,
\exp_{- \alpha, \beta - \alpha}(|\zeta|^{2}).
\label{F_val}
\eeq

\par

To conclude about the positivity of the measure, we, as it is   
evident from (\ref{F_intro}) and (\ref{F_val}), need to study the 
positivity of the 
deformed exponential $\eab ({\cal X})$ for arbitrary real arguments. 
We demonstrate this by adopting a method previously used in another 
context by Quesne. \ocite{Q04} The generalized exponential function 
$\eab ({\cal X})$ may be expressed as a product of ordinary 
exponentials: 
\beq
\eab ({\cal X}) = \exp (\sum_{k = 1}^{\infty}\,c_{k}\,{\cal X}^{k}),
\label{prod_exp}
\eeq
where the coefficients $c_{k}$ obey a linear recurrence relation
\beq
c_{k} = \frac{1}{(k)_{\alpha, \beta}!} - \frac{1}{k}\,
\sum_{\ell = 1}^{k - 1} \frac{\ell}{(k - \ell)_{\alpha, \beta}!}\,
c_{\ell}, \quad c_{1} = \frac{1}{(1)_{\alpha, \beta}}.
\label{rec_rel}
\eeq
The above triangular set of linear equations may be solved up to any 
arbitrary order, and the first few coefficients are written below:
\bea
&&c_{1} = \frac{1}{[\beta]_{q}},\qquad
c_{2} = \frac{1}{[\alpha]_{q}\, [\beta]_{q}\, [[\beta + \alpha/2]]}
- \frac{1}{2 [\beta]_{q}^{2}}, \nn\\
&&c_{3} = \frac{1}{[\alpha]_{q}\, [\beta]_{q}\, [\alpha + \beta]_{q}\,  
[[3 \alpha/2]]\,\, [[\beta + \alpha/2]]}
- \frac{1}{[\alpha]_{q}\, [\beta]_{q}^{2}\, [[\beta + \alpha/2]]}
+  \frac{1}{3 [\beta]_{q}^{3}}.
\label{rec_sol}
\eea
A consequence of the product structure (\ref{prod_exp}) is that the 
deformed exponential $\eab ({\cal X})$ is a positive definite quantity 
for real arguments. The above discussion leads us to infer that the 
measure function obtained via (\ref{res_unit}), 
(\ref{F_intro}), and (\ref{F_val}) 
\beq
\hbox{d}\mu(\zeta) = \hbox{d}^{2}\zeta \,
\frac{2 \alpha \,\, [[\beta - \alpha/2]]\,\, \ln q/(q - q^{-1})} 
{(\alpha\,\, [[\beta - \alpha/2]]\,\,\ln q/ (q - q^{-1}))^{2} + 
\varpi^{2}\,\,[\beta - \alpha/2]_{q}^{2}}\,\, \eab (|\zeta|^{2})\,\,
\exp_{- \alpha, \beta - \alpha}(|\zeta|^{2})
\label{mea_fin}
\eeq
is a positive definite quantity for $\alpha > 0$.
 
\section{Bipartite composite systems and entangled coherent states} 
\label{entanglement}
\setcounter{equation}{0}

The Hopf coalgebraic structure of the $\h$ algebra given in 
(\ref{U_coal}) is expected to play a qualitatively important role in 
describing the symmetry properties of many body systems. Keeping this 
picture in mind, we, in the present section, introduce the  normalized 
coherent states of the $\h$ algebra in the case of a bipartite 
composite system. The bipartite coherent states may be defined as  
\beq
\Delta(a) \, \z2 = \zeta \, \z2, \qquad \zeta \in {\mathbb C},
\label{bip_def}
\eeq 
where the noncocommutative coproduct structure $\Delta (a)$ is given in
(\ref{U_coal}}). Expanding of the state (\ref{bip_def}) in the tensored 
basis of the number states
\beq
\z2 = \sum_{n, m}\, c_{n, m}\, |n\rangle \otimes 
|m\rangle.
\label{z2_exp}
\eeq 
we obtain a double-indexed recurrence relations for the 
coefficients $c_{n, m}$:
\beq
c_{n+1, m}\,\sqrt{(n + 1)_{\alpha, \beta}}\,\, q^{(m \alpha + \beta)/2}
+ c_{n, m+1}\,(-1)^{n + \beta/\alpha}
\sqrt{(m + 1)_{\alpha, \beta}}\,\, q^{- (n \alpha + \beta)/2}
= \zeta \, c_{n, m}.
\label{2_rec}
\eeq
While describing the solution of the recurrence relation (\ref{2_rec}),
we, for the purpose of comparison, stay as close as possible to the
construction (\ref{coh_const}) of the coherent states of single-node 
systems. Pursuing this approach we consider the ansatz
\beq
c_{n, m} = \frac{\zeta_{1}^{n}\,\, \zeta_{2}^{m}}
{\sqrt{\nab!\,(m)_{\alpha, \beta}!}}\, g_{n, m},\qquad
(\zeta_{1}, \zeta_{2}) \in {\mathbb C}
\label{g_def}
\eeq
and redefine the parameters as follows
\beq
\frac{\zeta_{1}}{\zeta}\, q^{\beta/2} = \rho_{1},\qquad
(-1)^{\beta/\alpha}\,\frac{\zeta_{2}}{\zeta}\, q^{- \beta/2} = \rho_{2}
\label{rho_in}
\eeq
to obtain a simpler recurrence relation satisfied by the coefficients 
$g_{n, m}$:
\beq
\rho_{1}\, q^{m \alpha/2}\, g_{n+1, m}
+ (-1)^{n}\, \rho_{2}\, q^{- n \alpha/2}\, g_{n, m+1} = g_{n, m}.
\label{g_rec}
\eeq
We proceed towards solving the above recurrence relation by considering 
the coefficients $g_{n, m}$ as elements of a matrix. A little reflection 
then shows that given the elements of the first row we can obtain all 
other elements by employing (\ref{g_rec}) successively. Assuming the 
boundary condition 
\beq
g_{0, m}= d_{m},\qquad 0< d_{m} \leq 1
\label{ini_con}
\eeq
the solution of the recurrence relation (\ref{g_rec}) may be found by 
inspection. We quote the result: 
\bea
g_{n, m} &=& \frac{q^{- n m \alpha/ 2}}{\rho_{1}^{n}}\,
\sum_{k = 0}^{n}\,(- 1)^{k (k + 1)/ 2}\,
\left\lgroup\begin{array}{c}n\\k\end{array}
\right\rgroup_{q^{- \alpha}}\, \rho_{2}^{k}\, 
q^{- k (k -1) \alpha/2}\; d_{m + k},\nn\\
\big\lgroup\begin{array}{c}n\\k\end{array}
\big\rgroup_{q} &=& \frac{\big\lgroup n \big\rgroup_{q}!}
{\big\lgroup k \big\rgroup_{q}!\,\big\lgroup n - k \big\rgroup_{q}!},
\,\big\lgroup n \big\rgroup_{q}! = \prod_{\ell = 1}^{n}\,    
\big\lgroup \ell \big\rgroup_{q},  
\,\big\lgroup n \big\rgroup_{q} = \frac{1 - (-1)^{n}\,q^{n}}{1 + q}.
\label{g_sol_1}
\eea 
For special choices of the boundary coefficients $d_{m}$, the matrix 
elements $g_{n, m}$ may be expressed in closed form. For instance, 
choosing $d_{m} = \delta^{m},\, 0 < \delta \leq 1$, we obtain
\beq
g_{n, m} = q^{- n m \alpha/2}\,\frac{\delta^{m}}{\rho_{1}^{n}}\,
(\delta \,\rho_{2}; \, - q^{- \alpha})_{n},\qquad 
(a; q)_{n} = \prod_{\ell = 1}^{n}\, (1 - a\, q^{\ell - 1}).
\label{g_sol}
\eeq

\par

At this point it is useful to examine (\ref{g_sol}) in the 
$q \rightarrow 1$ limit. Retaining the previous notations, the 
coefficients $g_{n, m}$ in the $q \rightarrow 1$ limit are found 
to assume the form
\beq
g_{2 \ell, m} = \delta^{m}\,\left(\frac{1 - \delta^{2}\, 
\rho_{2}^{2}}{\rho_{1}^{2}}\right)^{\ell},\qquad
g_{2 \ell + 1, m} = \delta^{m}\,\frac{1 - \delta\,\rho_{2}}{\rho_{1}}
\left(\frac{1 - \delta^{2}\, \rho_{2}^{2}}{\rho_{1}^{2}}\right)^{\ell}.
\label{cla_sol}
\eeq 
It is apparent from (\ref{cla_sol}) that the coherent state of the 
composite bipartite system, in the $q \rightarrow 1$ limit, may be 
factorized in the states of the single-node subsystems. 

\par

Returning to the deformed case $(q \neq 1)$, we combine
(\ref{z2_exp}), (\ref{g_def}), (\ref{rho_in}) and (\ref{g_sol}) to finally 
obtain the bipartite coherent state as
\beq
\z2 = \sum_{n, m}\,q^{- n m \alpha/2}\,\frac{\delta^{m}\,\,
\zeta_{1}^{n}\,\, \zeta_{2}^{m}}{\rho_{1}^{n}\,\,
\sqrt{\nab!\,(m)_{\alpha, \beta}!}}\, 
(\delta \,\rho_{2}; \, - q^{- \alpha})_{n}
\,\,|n\rangle \otimes |m\rangle.
\label{z2_state}
\eeq
In the above construction of the bipartite coherent state $\z2$, the 
complex variables $\rho_{1}$ and $\rho_{2}$ (or, equivalently, 
$\zeta_{1}$ and $\zeta_{2}$) enter as arbitrary parameters. 
The norm of the bipartite coherent state (\ref{z2_state}) may now be
readily obtained as 
\beq
{\cal N} \equiv \langle \widehat{\zeta}\z2 = \sum_{n, m}|c_{n, m}|^{2}
= \sum_{n} \frac{q^{-n \beta}\,|\zeta|^{2 n}}{\nab !}\,\,
|(\delta \,\rho_{2}; \, - q^{- \alpha})_{n}|^{2}\,
\eab\Big(\delta^{2}\,|\zeta_{2}|^{2}\,q^{- n \alpha}\Big).
\label{z2_norm}
\eeq
Using the definition of the deformed exponential function 
(\ref{def_exp}) the norm has been expressed as a single sum. Its 
convergence in various domains may be tested in a straightforward way.
For instance, in the region $q > 1, (\alpha, \beta) > 0$, the sum is 
convergent, and, therefore, the norm (\ref{z2_norm}) is finite. The 
normalized coherent state ${\cal N}^{- 1/2}\,|\widehat{\zeta}\rangle$ 
of the bipartite system obeying the $\h$ Hopf symmetry may be 
readily obtained from (\ref{z2_state}) and (\ref{z2_norm}). The most 
remarkable property of the bipartite coherent state (\ref{z2_state}) 
is its naturally entangled structure for a nonclassical $(q \neq 1)$ 
value of the deformation parameter. {\it The summand in} (\ref{z2_state}) 
{\it include the factor $q^{- n m \alpha/2}$, which forbids factorization 
of the coherent state of the composite system into quantum states of 
single-node components}.

\section{Conclusion}
\label{Conclusion}
In conclusion we briefly mention the possibilities of further 
development of the topics discussed here. In connection with our 
derivation of the closed form expression of the universal ${\cal T}$
matrix capping the Hopf duality structure of the $\h$ and the $\H$
algebras, we mention the followings. Transfer matrices of integrable 
models are finite dimensional representations of the operator valued 
universal ${\cal T}$ matrix. As Calogero-Sutherland type of models 
are known \ocite{TPJ97} to have close kinship with the $\h$ algebra,
our universal ${\cal T}$ matrix may be of use in finding new 
deformations of these models. Moreover, $\h$ algebra is \ocite{MM92} 
the bosonized version of the Hopf superalgebra 
${\cal U}_{q}(osp(1|2))$. Our universal ${\cal T}$ matrix may provide 
direct clues on the derivation of as yet unknown universal ${\cal T}$ 
matrix of the ${\cal U}_{q}(osp(1|2))$ algebra. This is likely to be 
useful in constructing ${\cal U}_{q}(osp(1|2))$ based integrable models.

\par

Entanglement of states is the key feature in quantum information 
processing, such as quantum teleportation, \ocite{BBCJPW93} quantum
key distribution \ocite{E91} and so on. Our work raises the interesting 
possibility that the composite systems such as anyons \ocite{CGM93}
for instance, may be naturally equipped for implementing entangled 
states. If the composite system allows for a variation of the deformation 
parameter $q$, a `switching mechanism' for entanglement may be 
developed. Lastly, the multipartite systems subject to $\h$ symmetry    
show new levels of entanglement. We will return to this topic in a future 
work. 

%
\section*{Acknowledgements}

  The work of N.A. is partially supported by the grants-in-aid 
from JSPS, Japan (Contract No. 15540132). The other authors (R.C. and 
J.S.) are partially supported by the grant DAE/2001/37/12/BRNS, Government 
of India.


\begin{thebibliography} 
{99} 

\bibitem{M89}
{A.J. Macfarlane, J. Phys. {\bf A22} (1989) 4581.}

\bibitem{B89}
{L.C. Biedenharn,  J. Phys. {\bf A22} (1989) L873.}

\bibitem{SF89}
{C.P. Sun and H.C. Fu, J. Phys. {\bf A22} (1989) L983.}

\bibitem{CJ91}
{R. Chakrabarti and R. Jagannathan, J. Phys. {\bf A24} (1991) L711.}

\bibitem{V91}
{M.A. Vasiliev, Int. J. Mod. Phys. {\bf A6} (1991) 1115.}

\bibitem{BEM93}
{T. Brezizinski, I.L. Egusquiza and A.J. Macfarlane, Phys. Lett.
{\bf B311} (1993) 202.}

\bibitem{M94}
{A.J. Macfarlane, J. Math. Phys. {\bf 35} (1994) 1054.}

\bibitem{CJ94}
{R. Chakrabarti and R. Jagannathan, J. Phys. {\bf A27} (1994) L277.}

\bibitem{Y91}
{H. Yan, Phys. Lett. {\bf B262} (1991) 459.}
 
\bibitem{MM92} 
{A.J. Macfarlane and S. Majid, Int. J. Mod. Phys. {\bf A7} (1992) 
4377.}

\bibitem{OS94}
{C.H. Oh and K. Singh, J. Phys. {\bf A27} (1994) 5907.}
 
\bibitem{PT97} 
{A. Paolucci, I. Tsohantjis, Phys. Lett. {\bf A234} (1997) 27.}

\bibitem{TPJ97}
{I. Tsohantjis, A. Paolucci and P.D. Jarvis, J. Phys. {\bf A30}
(1997) 4075.}

\bibitem{MT97}
{D.S. McAnally and I. Tsohantjis, J. Phys. {\bf A30} (1997) 651.}

\bibitem{APS01}
{G. Alexanian, A. Pinzul and A. Stern, Nucl. Phys. {\bf B600}
(2001) 531.}

\bibitem{C69}
{F. Calogero, J. Math. Phys. {\bf 10} (1969) 2191.}
 
\bibitem{S71}
{B. Sutherland, J. Math. Phys. {\bf 12} (1971) 246.}
 
\bibitem{FG93}
{C. Fronsdal and A. Galindo, Lett. Math. Phys. {\bf 27} (1993) 59.} 

\bibitem{SPS99}
{J.M. Sixdeniers, K.A. Penson and A.I. Solomon, J. Phys. {\bf A32}
(1999) 7543.}

\bibitem{KPS01}
{J.R. Klauder, K.A. Penson and J.M. Sixdeniers, Phys. Rev. 
{\bf A64} (2001) 013817.}

\bibitem{Q02}
{C. Quesne, J. Phys. {\bf A35} (2002) 9213.}

\bibitem{CV04}
{R. Chakrabarti and S.S. Vasan, J. Phys. {\bf A37} (2004) 10561.}

\bibitem{S92}
{B.C. Sanders, Phys. Rev. {\bf A45} (1992) 6811.}

\bibitem{MMS00}
{W.J. Munro, G.J. Milburn, B.C. Sanders, Phys. Rev. {\bf A62} (2000) 
052108.} 

\bibitem{BG71}
{A.O. Barut, L. Girardello, Comm. Math. Phys. {\bf 21} (1971) 41.}

\bibitem{CV03}
{R. Chakrabarti and S.S. Vasan, Phys. Lett. {\bf A312} (2003) 287.} 

\bibitem{Q04}
{C. Quesne, Int. J. Theor. Phys. {\bf 43} (2004) 545.}  

\bibitem{BBCJPW93}
{C.H. Bennet, G. Brassard, C. Crepeau, R. Josza and W.K. Wooters,
Phys. Rev. Lett. {\bf 70} (1993) 1895.}

\bibitem{E91}
{A.E. Ekert, Phys. Rev. Lett. {\bf 67} (1991) 661.}

\bibitem{CGM93}
{M. Chaichian, R. Gonzales Felipe and C. Montonen,
J. Phys. {\bf A26} (1993) L1117.} 
\end{thebibliography}
\end{document}